\documentclass[a4paper,fleqn]{cas-dc}

\usepackage[authoryear,longnamesfirst]{natbib}
\usepackage{lineno,hyperref}
\usepackage{ulem}
\usepackage{algorithm2e,setspace}
\usepackage{subfigure}
\usepackage{multirow}
\usepackage{bbding}
\usepackage{mathrsfs}
\usepackage{amsthm,amsmath,amssymb}
\usepackage{tabularx}
\usepackage{natbib}
\usepackage{threeparttable}

\def\tsc#1{\csdef{#1}{\textsc{\lowercase{#1}}\xspace}}
\tsc{WGM}
\tsc{QE}
\tsc{EP}
\tsc{PMS}
\tsc{BEC}
\tsc{DE}
\begin{document}
\let\WriteBookmarks\relax
\def\floatpagepagefraction{1}
\def\textpagefraction{.001}

\shorttitle{}

\shortauthors{Xie et~al.}

\title [mode = title]{Adversarial multi-task underwater acoustic target recognition: towards robustness against various influential factors}                      
% \tnotemark[1,2]


% \tnotetext[1]{Corresponding author at: Institute of Acoustics, Chinese Academy of Sciences, No. 21, Beisihuan West Road, Haidian District, Beijing, China.}

% \tnotetext[1]{E-mail addresses: xieyuan@hccl.ioa.cn.cn (Yuan Xie), renjiawei@hccl.ioa.ac.cn (Jiawei Ren), xuji@hccl.ioa.ac.cn (Ji Xu).}

\author[address1,address2]{Yuan Xie}[%type=editor,
                        %auid=000,bioid=1,
                        %prefix=Sir,
                        %role=Researcher,
                        style=chinese,
                        orcid=0000-0003-3803-0929]

%\cormark[1]

% \fntext[fn1]{First author.}
\ead{xieyuan@hccl.ioa.cn.cn}

%\ead{cvr_1@tug.org.in}

%\ead[url]{www.cvr.cc, cvr@sayahna.org}

\credit{Conceptualization, Methodology, Software, Validation, Formal analysis, Investigation, Data Curation, Writing - Original Draft, Writing - Review \& Editing, Visualization}

\author[address1,address2,address3]{Ji Xu}[style=chinese]
\ead{xuji@hccl.ioa.cn.cn}
\credit{Data Curation, Supervision, Resources}

\author[address1,address2]{Jiawei Ren}[style=chinese]
\ead{renjiawei@hccl.ioa.cn.cn}
\credit{Methodology, Investigation}

\author[address1,address2]{Junfeng Li}[style=chinese]
\ead{lijunfeng@hccl.ioa.cn.cn}
\credit{Writing - Review \& Editing, Supervision, Project administration, Funding acquisition}
\cormark[1]


\affiliation[address1]{organization={Key Laboratory of Speech Acoustics and Content Understanding, Institute of Acoustics, Chinese Academy of Sciences},
    addressline={No.21, Beisihuan West Road, Haidian District},
    postcode={100190},
    city={Beijing},
    country={China}}
    
\affiliation[address2]{organization={University of Chinese Academy of Sciences},
    addressline={No.80, Zhongguancun East Road, Haidian District}, 
    postcode={100190},
    city={Beijing},
    country={China}}

\affiliation[address3]{organization={State Key Laboratory of Acoustics, Institute of Acoustics, Chinese Academy of Sciences},
    addressline={No.21, Beisihuan West Road, Haidian District},
    postcode={100190},
    city={Beijing},
    country={China}}

\cortext[cor1]{Corresponding author}
%\cortext[cor2]{Principal corresponding author}

% \fntext[fn1]{The source code of this paper could be obtained from https://github.com/xy980523/UATR-CMoE.}
% \fntext[fn2]{Another author footnote, this is a very long footnote and
%   it should be a really long footnote. But this footnote is not yet
%   sufficiently long enough to make two lines of footnote text.}

% \nonumnote{This note has no numbers. In this work we demonstrate $a_b$
%   the formation Y\_1 of a new type of polariton on the interface
%   between a cuprous oxide slab and a polystyrene micro-sphere placed

\begin{abstract}
Underwater acoustic target recognition based on passive sonar faces numerous challenges in practical maritime applications. One of the main challenges lies in the susceptibility of signal characteristics to diverse environmental conditions and data acquisition configurations, which can lead to instability in recognition systems. While significant efforts have been dedicated to addressing these influential factors in other domains of underwater acoustics, they are often neglected in the field of underwater acoustic target recognition. To overcome this limitation, this study designs auxiliary tasks that model influential factors (e.g., source range, water column depth, or wind speed) based on available annotations and adopts a multi-task framework to connect these factors to the recognition task. Furthermore, we integrate an adversarial learning mechanism into the multi-task framework to prompt the model to extract representations that are robust against influential factors. Through extensive experiments and analyses on the ShipsEar dataset, our proposed adversarial multi-task model demonstrates its capacity to effectively model the influential factors and achieve state-of-the-art performance on the 12-class recognition task.
\end{abstract}



% \begin{highlights}
% \item Establishing transition from single-task learning to multi-task learning.
% \item Simple auxiliary task that enhances underwater target recognition with limited data.
% \item First work to apply multi-gate mixture-of-experts to the underwater acoustic domain.
% \item Optimized system by using frequency-domain features as input to the gating module.
% \item Remarkable advancements showcased on ShipsEar dataset, reaching state-of-the-art.
% \end{highlights}

% \begin{keyword}
% %% keywords here, in the form: keyword \sep keyword
% Underwater acoustic target recognition \sep ship-radiated noise \sep deep learning \sep multi-task learning \sep multi-gate mixture-of-experts
% %% PACS codes here, in the form: \PACS code \sep code
% %\PACS 0000 \sep 1111
% %% MSC codes here, in the form: \MSC code \sep code
% %% or \MSC[2008] code \sep code (2000 is the default)
% %\MSC 0000 \sep 1111
% \end{keyword}

\maketitle


\section{Introduction}
Underwater acoustic target recognition based on passive sonar plays a vital role in the field of marine acoustics~\cite{rajagopal1990target}. The primary objective of this task is to automatically classify the category of targets by analyzing the radiated noise signals received by passive sonar. The long detection range, reassuring concealment, and low deployment cost of passive underwater acoustic target recognition make it indispensable in practical applications~\cite{heupel2006automated}. This technology finds extensive applications in various areas, including underwater surveillance~\cite{sutin2010stevens}, marine resources development and protection~\cite{vaccaro1998past}, and security defense~\cite{fillinger2010towards}.

\begin{figure*}
    \centering
    \includegraphics[width=\linewidth]{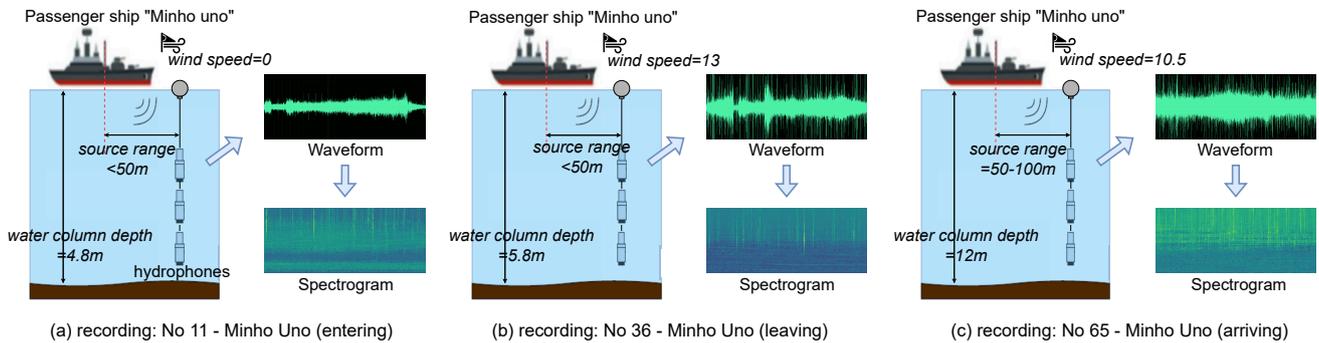}
    \caption{A comparison of signal waveforms and spectrograms belonging to the same target (a passenger ship ``Minho Uno'') under different influential factors. The samples are drawn from the ShipsEar dataset.}
    \label{fig:background}
    \vspace{-2px}
\end{figure*}

In recent years, data-driven machine learning techniques, particularly deep neural networks, have emerged as the dominant approaches for underwater acoustic recognition systems~\cite{xie2022underwater,sun2022underwater,xu2023underwater}. These techniques rely less on prior knowledge and manual parameter setting~\cite{ren2022ualf} and possess powerful modeling capabilities for diverse distributions of underwater acoustic data~\cite{irfan2021deepship}. However, in real-world application scenarios, machine learning-based recognition models often struggle to capture robust target-relevant characteristics due to the inherent instability and susceptibility of underwater acoustic signals to various environmental conditions and data acquisition configurations~\cite{santos2016shipsear,chen2021environment,xie2022underwater}. Environmental conditions, such as ambient noise, wind speed, water column depth, etc., can influence the amplitude and frequency components of signals through the generation of interference waves or by affecting the propagation of sound waves. Additionally, data acquisition configurations, such as the sound source range, the frequency response characteristics of receivers, etc., can also impact the power level and frequency distribution of signals. Fig.~\ref{fig:background} compares the waveforms and spectrograms of signals belonging to the same target (a passenger ship named ``Minho uno'') under different conditions of source range, water column depth, and wind speed. As shown in Fig.~\ref{fig:background} (b) and (c), when the wind speed reaches 10.5 or 13~km/h, strong winds can generate surface waves that propagate into the water column. These wind-induced surface waves have the potential to cause noticeable interference to signals, which can be observed as vertical stripes in the waveform and spectrogram. Furthermore, when the source range and water column depth increase, there is a noticeable attenuation of the high-frequency components of the signal (indicated by lighter colors in the spectrogram, representing lower energy). This propagation-induced attenuation can also affect the frequency distribution.

% Consequently, this limitation hampers the accuracy and generalization capability of underwater acoustic recognition systems. To address this issue, it is imperative to consider the impact of environmental conditions and enhance the robustness of recognition systems against variable environmental conditions.

According to our research, modeling environmental conditions and data acquisition configurations (referred to as ``influential factors'' in subsequent paragraphs) has been widely applied across various domains in underwater acoustics, including underwater acoustic networks~\cite{jiang2008underwater,domingo2008overview}, underwater acoustic communications~\cite{yang2012properties,khan2020channel,chen2021environment}, underwater source localization~\cite{hassab1984influence,worthmann2017adaptive}, and others. However, in the domain of underwater acoustic target recognition, which primarily relies on data-driven machine learning, the limited availability of data and related annotations concerning influential factors hinders the research development. While partial researchers have acknowledged the influence of these factors in signal characteristics~\cite{xie2022underwater,li2023data}, there remains a gap in establishing a clear association between these factors and the recognition task. To address the issue, previous recognition-related studies have primarily focused on mitigating marine environment noise  ~\cite{zhou2020denoising,zhou2023novel} or generating synthetic signals to simulate data under different influential factors for data augmentation~\cite{li2023data}. However, these approaches may yield limited improvements or even lead to a decline in performance due to the discrepancies between manually estimated conditions and the actual underwater environment~\cite{xu2023underwater}.

To address this research gap, this study adopts a multi-task framework to leverage influential factors in a direct manner. We selected several influential factors (source range, water column depth, wind speed) with relatively complete annotations from the ShipsEar dataset~\cite{santos2016shipsear}, and designed relevant auxiliary task to model specific influential factors. The auxiliary task can facilitate the transfer of knowledge about influential factors to the recognition task through shared network layers and parameters. Furthermore, we integrate an extra adversarial learning mechanism into the multi-task framework. This mechanism can discourage the model from focusing on discriminative patterns related to the influential factors, thereby prompting the front-end shared layer of the model to extract robust representations that are insensitive to influential factors. In addition, we also introduce two customized optimization techniques, which serve to address potential issues of our proposed adversarial multi-task model. To evaluate the efficacy of our proposed model, we conducted extensive experiments on the ShipsEar dataset, demonstrating substantial performance enhancements compared to baselines and current advanced methods. Our AMTNet can achieve state-of-the-art results on the 12-class recognition task, with an average accuracy of 80.95\%. The key contributions of this study can be summarized as follows:

\begin{itemize}
\item This study identifies the limitations of current recognition models, which are susceptible to environmental conditions and data acquisition configurations, and pioneers the modeling of these influential factors in the underwater acoustic recognition domain;

\item We design auxiliary tasks that model influential factors based on available annotations and adopt a multi-task framework as a bridge to connect these factors with the recognition task;

\item We incorporate an adversarial learning mechanism into the multi-task framework, aiming to enhance the model's robustness against influential factors.

% \item Our proposed model achieves state-of-the-art performance on the ShipsEar's 12-class recognition task, with an accuracy of 80.95\%;

\end{itemize}

\section{Related Works}
With the advancement of computing resources and the availability of open-source underwater acoustic databases~\cite{irfan2021deepship,santos2016shipsear}, data-driven machine learning methods have emerged as the predominant approach for underwater acoustic target recognition. In particular, over the past five years, there has been a growing interest in applying deep neural networks for underwater recognition tasks. As reported in the literature, Zhang et al.~\cite{zhang2021integrated} utilized the short-time Fourier transform (STFT) amplitude spectrum, STFT phase spectrum, and bispectrum features as inputs for convolutional neural networks (CNNs); Liu et al.~\cite{liu2021underwater} employed convolutional recurrent neural networks with 3-D Mel-spectrograms and data augmentation for underwater target recognition; Xie et al.~\cite{xie2022adaptive} utilized learnable fine-grained wavelet spectrograms with the deep residual network (ResNet)~\cite{he2016deep} to adaptively recognize ship-radiated noise; Ren et al.~\cite{ren2022ualf} employed learnable Gabor filters and ResNet for constructing an intelligent underwater acoustic classification system, etc. In contrast to classical recognition paradigms~\cite{das2013marine,wang2014robust}, models based on deep neural networks possess the ability to leverage a vast number of parameters and complicated nonlinear operators, enabling complex modeling and facilitating the extraction of high-level patterns from intricate underwater signals.

Despite the promising performance demonstrated on current open-source underwater databases, models based on deep neural networks encounter limitations in their generalization capabilities when applied to real-world underwater scenarios~\cite{xie2022underwater}. These data-driven approaches can develop biased perceptions of the marine environment and target-relevant characteristic patterns in cases where the available training data is insufficient. This may lead to the overfitting of recognition models~\cite{li2023data,xu2023underwater}, rendering them sensitive to influential factors in real-world underwater scenarios, thereby posing a challenge to realize robust recognition. To mitigate the systems' susceptibility to influential factors and enhance their generalization capabilities for unseen data, various studies in other domains of underwater acoustics have made efforts to model these influential factors, aiming to mitigate their interference. For instance, Domingo, M. C.~\cite{domingo2008overview} presented several channel models concerning fading, multipath, and refractive properties of the sound channel for underwater communication networks; Chen et al.~\cite{chen2021environment} proposed environment-aware communication channel quality prediction for underwater acoustic transmissions; Hassab, J.~\cite{hassab1984influence} analyzed the influence of unequal sub-array spacing configurations on source localization, and so on. However, in the field of underwater acoustic target recognition, only a limited number of studies have taken into account the influential factors. Some of these studies employed environmental noise reduction techniques to mitigate the interference caused by the surroundings~\cite{zhou2020denoising,zhou2023novel}, while others have focused on data augmentation by simulating synthetic signals under specific environmental conditions~\cite{li2023data}. These approaches, which aim to capture the data distribution in real-world scenarios, may not always yield consistent improvements~\cite{xu2023underwater} and can sometimes lead to performance degradation due to the inevitable deviations~\cite{gong2021eliminate} between manual simulation and real-world underwater environment.

To address the inadequacies of current recognition techniques, it is crucial to directly model influential factors without relying on manual simulation. Multi-task learning, which has exhibited significant success in many areas~\cite{ma2018modeling,tang2020progressive}, offers a promising solution. By considering the estimation of influential factors as the auxiliary task, the multi-task model can acquire knowledge related to these factors, thereby improving its ability to comprehensively perceive underwater signals. Moreover, studies in speech recognition~\cite{shinohara2016adversarial,adi2019reverse}, speech enhancement~\cite{du2020self,qiu2022adversarial}, and other domains~\cite{liu2017adversarial,mao2020adaptive,liang2020aspect} have explored the integration of multi-task learning with adversarial learning techniques, which can be referred to as adversarial multi-task learning, to explicitly enhance the invariance and robustness of representations shared across multiple tasks. In the domain of underwater acoustic target recognition, while several studies have utilized multi-task learning to assist in feature extraction~\cite{li2023robust} or autoencoder training~\cite{lei2023deep}, the application of multi-task learning specifically focused on influential factors remains unexplored. Besides, to the best of our knowledge, adversarial multi-task learning has not yet been applied in this field.

%In the field of underwater acoustic recognition, several works have employed multi-task learning techniques. However, these approaches have not taken influential factors into account. Furthermore, to the best of our knowledge, there is currently no utilization of adversarial training for modeling influential factors in underwater acoustic recognition. Thus, this work can be considered the first proposal for adversarial multi-task learning to achieve robust recognition by incorporating influential factors in this field.}

% In this study, we refrain from modifying the training data and instead introduce additional annotations regarding environment-related factors to encourage the model to gain a deeper and more comprehensive understanding of the data, thereby reducing its sensitivity to complex environments in a risk-free manner.

\section{Methods}

This section is organized as follows. We first describe the pre-processing and feature extraction processes in subsection A. Next, in subsection B, we introduce the design of the auxiliary tasks related to three candidate influential factors. Then, in subsection C, we present the overall training and testing procedures of our proposed adversarial multi-task model. Following that, in subsection D, we provide a detailed description of the model architecture for each module. Finally, we introduce two customized optimization techniques in subsection E.

\subsection{Pre-Processing and Feature Extraction}

\begin{figure*}
    \centering
    \includegraphics[width=\linewidth]{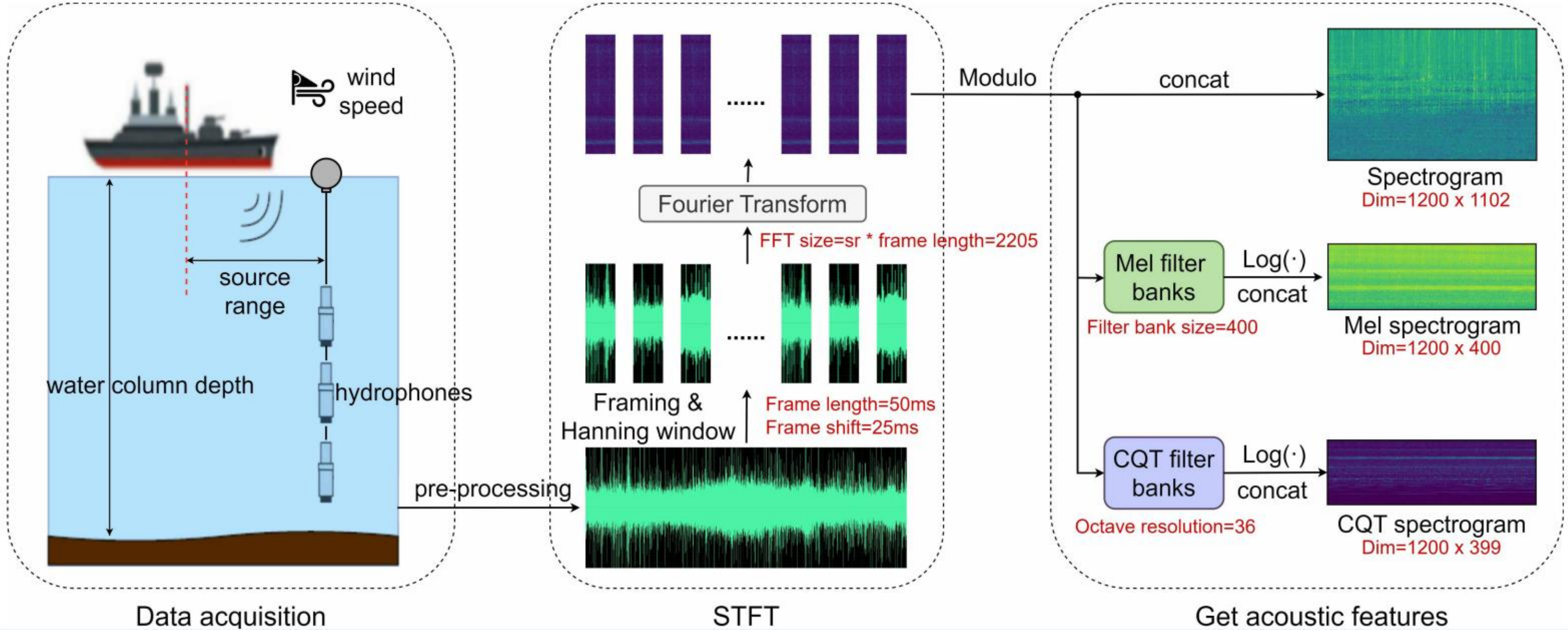}
    \caption{The detailed process of data acquisition and feature extraction. Parameter setups or feature dimensions are displayed in red characters.}
    \label{fig:fea_extract}
    \vspace{-2px}
\end{figure*}

The detailed process of pre-processing and feature extraction is illustrated in Fig.~\ref{fig:fea_extract}. Initially, ship-radiated noise signals were collected using single or multiple hydrophones. In cases where multiple hydrophones were employed, the recording with the highest sound level was selected for further analysis. For this study, we conducted experiments on the open-source dataset~\cite{santos2016shipsear}, which grants direct access to collected single-channel signals. The signals were then resampled to 44100~Hz and filtered using a band-pass filter with a cutoff frequency of 10~Hz-22050~Hz to eliminate noise interference. The selection of the cutoff frequency was determined through relevant preliminary experiments, as detailed in Section V.A. Following this, a mean-variance normalization was applied to standardize the waveform amplitude ranges, ensuring they fall within an interval with a mean of 0 and a variance of 1.

After the pre-processing stage, feature extraction was performed on the normalized signals. In this study, we employed three acoustic features that have demonstrated promising performance in prior research: spectrogram~\cite{xu2023underwater}, Mel spectrogram~\cite{liu2021underwater}, and CQT spectrogram~\cite{irfan2021deepship}. Examples of these features, along with their corresponding feature dimensions, are displayed in Fig.~\ref{fig:fea_extract}. To extract these features, the normalized signals were first framed using a 50~ms Hanning window with 50\% overlap. Afterward, the short-time Fourier transform (STFT) was applied to the windowed frames to derive the complex spectra, whose FFT size was determined as the number of sampling points included in one frame (FFT size=44100 Hz$\times$0.05 second=2205). Each complex spectrum represents the local frequency content of the framed signal at different time points. Then:

\textbf{Spectrogram:} the complex spectra were transformed into amplitude spectra through a modulus operation. These amplitude spectra were then concatenated across all frames to obtain the spectrogram.

\textbf{Mel spectrogram:} the amplitude spectra were passed through Mel filter banks, which consist of a set of triangular bandpass filters spaced according to the non-linear Mel scale: 

\begin{equation}
    Mel(f)=2595\times log(1+\frac{f}{700}),
\end{equation}

% Each filter is centered at a specific Mel frequency and has a bandwidth that varies based on the non-linear Mel scale. 

where the number of filter banks was set to 400. The spectra mapped to the Mel scale were then logarithmically transformed and concatenated across all frames to derive the Mel spectrogram. The Mel spectrogram provides higher frequency resolution at low frequencies, facilitating the analysis of low-frequency line spectrum components.

\textbf{CQT spectrogram:} the amplitude spectra underwent the constant-Q transform (CQT), which convolved the spectra with a bank of bandpass filters (CQT kernel) that are logarithmically spaced in frequency. Among filter banks, the center frequency component of the $k$-th filter, denoted as $f_k$, was determined as:

\begin{equation}
    CQT(f_k)=2^{k/b} f_{min}, \quad k=0,1,\ldots,\lceil b\cdot log_2(\frac{f_{max}}{f_{min}}) \rceil -1
\end{equation}

where the upper and lower frequencies to be processed are represented by $f_{max}$=22050~Hz and $f_{min}$=10~Hz respectively, and the octave resolution is denoted by $b$ (typically $b$=36). The ratio of the filter bandwidth $BW$ to the center frequency is a constant $Q=\frac{f_k}{BW}=\frac{1}{2^{1/b}-1}$. The filtered spectra were then logarithmically transformed and concatenated across all frames to derive the CQT spectrogram. The CQT spectrogram offers higher frequency resolution in the low-frequency range and higher temporal resolution in the high-frequency range. It not only facilitates the analysis of low-frequency components but also provides information on periodic modulation in the high-frequency portion, such as propeller rhythm.

\subsection{Design of Auxiliary Tasks}

Compared to conventional recognition methods, our proposed method makes a breakthrough by integrating auxiliary tasks to model influential factors. It enables the transfer of knowledge regarding influential factors to the recognition task through parameter sharing, leading to improved perceptions of signal characteristics. In this study, we leveraged available metadata (e.g., acquisition location and date, source range, water column depth, atmospheric and oceanographic data, wind speed) from the ShipsEar dataset. The annotations for metadata were directly recorded during the data acquisition process, rather than being estimated through signal processing methods. Consequently, these annotations can be deemed reliable and accurate. Among the available metadata, we selected three factors that possess sufficient annotations and exhibit apparent impacts on underwater signals to construct the auxiliary tasks:

\textbf{Source range} refers to the distance between the target and the vertical of the hydrophone (see ``source range'' in Fig.~\ref{fig:fea_extract}). It influences the intensity or amplitude of underwater signals. Additionally, various frequency components may also be affected to varying degrees by range-dependent effects such as attenuation, scattering, and reverberation patterns;

\textbf{Water column depth} refers to the vertical depth between the water surface and the seafloor (see ``water column depth'' in Fig.~\ref{fig:fea_extract}). It influences the propagation path of sound waves and their interactions, thereby affecting the arrival time, amplitude, and frequency content of received signals. In shallow water columns, the seafloor can significantly influence the received signals through reflection, scattering, and absorption;

\textbf{Wind speed} refers to the speed at which air moves horizontally across the water surface. It affects the generation of wind-induced surface waves, which introduce surface noise that can propagate into the water column and interfere with underwater signals. Besides, the agitation of the water surface by wind generates turbulence and wave breaking, giving rise to broadband ambient noise.

\begin{figure*}[htbp]
    \centering
    \includegraphics[width=0.8\linewidth]{figs-amt/Figure3.pdf}
    \caption{The distribution of different categories of data in the ShipsEar dataset under various influential factors: (a) source ranges; (b) water column depths; (c) wind speeds.}
    \label{fig:distribution}
    \vspace{-2px}
\end{figure*}

%Considering the limited number of signal recordings, directly estimating the annotated environmental conditions is difficult due to the lack of sufficient samples. There is only one record under some environmental conditions, which is not enough to support the training test division to verify the performance of auxiliary tasks. 

Based on these three influential factors, we consider the estimation of the influential factor as the objective of the auxiliary task. It is worth mentioning that each auxiliary task is designed to individually estimate a specific influential factor, rather than attempting to predict all influential factors simultaneously. This design allows for a more targeted and accurate estimation of each factor.

Subsequently, we shall take the data circumstances into account and further refine the auxiliary tasks. The distribution of different categories of data in the ShipsEar dataset under various influential factors is illustrated in Fig.~\ref{fig:distribution}. It can be observed that the limited number of annotated recordings in ShipsEar and their unbalanced distribution pose challenges for estimating influential factors on a fine-grained level. For instance, there is only one signal recording with a source range exceeding 200 meters, which is inadequate to support model training and validation. To facilitate the training of auxiliary tasks, we established a label mapping standard to group recordings with various influential factors into specific classes based on defined criteria~\cite{matthews1993ramsar,pacheco2015retrieval,leu2005remotely,wheeler2004calm}. It ensures an adequate number of samples in each class to support the training of auxiliary tasks, with each class assigned a one-hot code as its label. The specific label mapping standard, along with corresponding specifications and the number of recordings for each class, is presented in Table~\ref{tab:aux}.

\begin{table*}[ht]
\normalsize
    \caption{\label{tab:aux} The label mapping standard, including the grouping thresholds, corresponding labels, specifications, and the number of recordings.}
    \centering
	\scalebox{0.65}{
	\begin{tabular}{lllc}
        \hline
	Influential Factor\quad & Label Mapping&Specifications\quad&Num\\
	\hline
	Source range (s)&Label=0 if 0$<$s$<$50 meters&Close source range&65\\
        & Label=1 if 50$\leq$ s$\leq$ 350 meters&Medium source range&25\\
        %&&2 -- ``more than 100 meters'',``350 meters''&3\\
        \hline
        Water column depth (d)&Label=0 if 0$<$d$<$6 meters&Land-sea interface or coastal wetland defined by Ramsar Convention&33\\
        &Label=1 if 6$\leq$ d$\leq$ 12 meters&Coastal zones, shallow water defined by~\cite{pacheco2015retrieval}&33\\
        &Label=2 if 12$<$ d$\leq$ 20 meters&Deep water region defined by~\cite{leu2005remotely}&24\\
        \hline
        Wind speed (w) &Label=0 if w=0 km/h&``Calm'' defined by Beaufort Wind Scale (BWS), sea like a mirror&23\\

        &Label=1 if 0$<$w$<$11 km/h&``Light Air/Breeze'' defined by BWS, small wavelets on sea&29\\
        &Label=2 if 11$\leq$w$\leq$18 km/h&	``Gentle Breeze'' defined by BWS, large wavelets on sea&25\\
        & ``Not available''&&13\\
        \hline
	\end{tabular}}
\end{table*}

Regarding the source range, since the majority of samples (72\%) have a close source range of less than 50~meters, we combined the remaining samples with a medium source range (50$\sim$350~meters) into another class. This approach ensures that the number of samples in each class does not differ significantly, thereby converting the source range estimation into a 2-class classification task. As for the water column depth, we established the threshold of 6~meters and 12~meters based on the definitions provided by the Ramsar Convention~\cite{matthews1993ramsar} and previous studies~\cite{pacheco2015retrieval,leu2005remotely}, respectively. The samples are divided into three classes: land-sea interface, shallow water, and deep water. In this way, water column depth estimation is converted into a 3-class classification task. Regarding the wind speed, we utilized the Beaufort Wind Scale (BWS)~\cite{wheeler2004calm} and set the threshold of 0~km/h and 11~km/h. The samples are then grouped into three classes: calm, light air/breeze, and gentle breeze. Thus, the wind speed estimation can be treated as a 3-class classification task. Notably, we encountered 13 samples lacking annotations for wind speed, and therefore, these samples were excluded from the auxiliary task training. Table~\ref{tab:aux} also presents the number of recordings for each class, demonstrating a relatively balanced distribution across different classes.

\subsection{Detailed Process of Model Training and Testing}

\begin{figure*}
    \centering
    \includegraphics[width=0.85\linewidth]{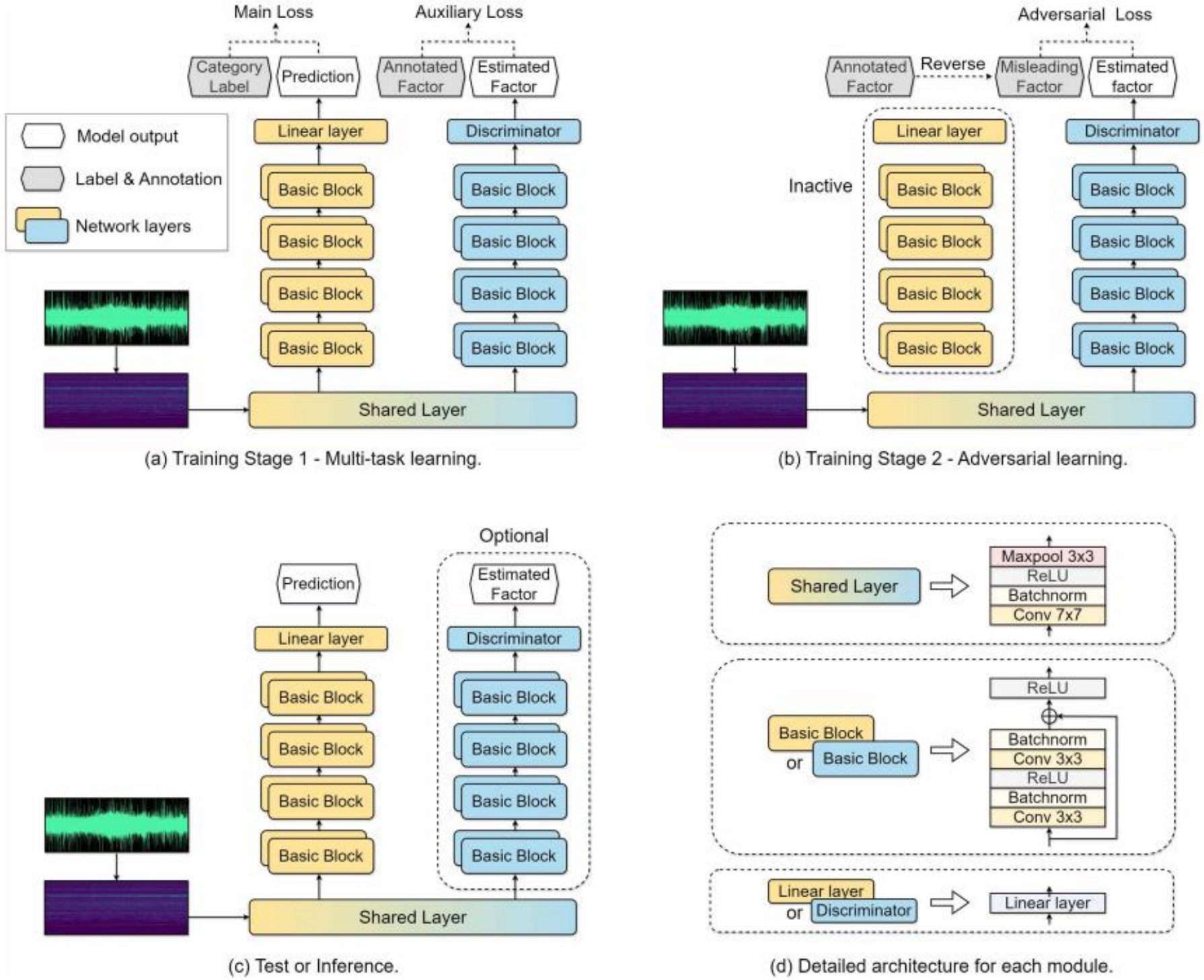}
    \caption{The detailed process of the training and test stage of our proposed AMT.}
    \label{fig:train}
    \vspace{-2px}
\end{figure*}

After performing feature extraction and setting up auxiliary tasks, we employ a multi-task framework along with adversarial learning to construct a model, called AMTNet. The detailed training and testing procedures of AMTNet are illustrated in Fig.~\ref{fig:train}. The overall framework of AMTNet consists of three primary components: the recognition branch responsible for predicting the target category (shown in yellow in Fig.~\ref{fig:train}); the auxiliary branch utilized to estimate influential factors (shown in blue in Fig.~\ref{fig:train}); and the shared layer responsible for acquiring general representations for both tasks and facilitating communication between the two branches to transfer knowledge (shown in yellow-blue gradient in Fig.~\ref{fig:train}).

%%%%%%%%% HERE

% \begin{table}[ht]
% \normalsize
%     \caption{\label{tab:aux} Auxiliary tasks and their corresponding label mapping relation.}
%     \centering
% 	\scalebox{1.0}{
% 	\begin{tabular}{ll}
%         \hline
% 	Task&Label mapping (Label -- Annotations)\\
% 	\hline
% 	source range estimation&0 -- ``less than 50 meters''\\
%         &1 -- ``50-100 meters'',``100 meters''\\
%         &2 -- ``more than 100 meters'',``350 meters''\\
%         \hline
%         water column depth estimation &0 -- ``4.8 meters'',``5.8 meters''\\
%         &1 -- ``7.5 meters'',``9.1 meters'',``12 meters''\\
%         &2 -- ``14 meters'',``15 meters'',``20 meters''\\
%         \hline
%         Wind speed estimation &0 -- ``0'',``1.93'',``2'',``3.34'',``5.29''\\
%         &1 -- ``10.5'',``12.21'',``13'',``18''\\
%         &2 -- ``Not available''\\
%         \hline
% 	\end{tabular}}
% \end{table}

Next, we provide a detailed exposition of the two-stage training process of AMTNet. Denote the input features as $X=\{x_1,x_2,...,x_N\}$, where $N$ represents the batch size. Each sample $x_i (i=1,2,...,N)$ is associated with a category label $y_i$ and an auxiliary label $\widetilde{y_i}$ regarding the influential factor. In the first stage, known as the multi-task learning stage (see Fig.~\ref{fig:train}(a)), the features $X$ are fed into the shared layer to obtain the general representations $R=\{r_1,r_2,...,r_N\}$. These representations $R$ are then simultaneously fed into the recognition branch and the auxiliary branch. The recognition branch consists of a module $F_{\mathrm{main}}(\cdot)$, composed of several sequential basic blocks, and a fully connected layer $fc(\cdot)$. In parallel, the auxiliary branch comprises a module $F_{\mathrm{aux}}(\cdot)$, which shares the same structure as $F_{\mathrm{main}}(\cdot)$ but has distinct parameters, accompanied by a linear discriminator $dis(\cdot)$. The recognition branch outputs the predicted probabilities $P=\{p_1,p_2,...,p_N\}\sim \mathcal{R}^{N\times n_{\mathrm{class}}}$ for target categories, while the auxiliary branch outputs the estimated probabilities $\widetilde{P}=\{\widetilde{p}_1,\widetilde{p}_2,...,\widetilde{p}_N\}\sim \mathcal{R}^{N\times n_{\mathrm{aux}}}$ for influential factors. Here, $n_\mathrm{class}$ represents the number of target categories (in this study, $n_\mathrm{class}$=12), while $n_{\mathrm{aux}}$ represents the number of mapped classes of influential factors ($n_\mathrm{class}$=2 or 3). The outputs of the two branches can be formulated as:

\begin{equation}
\begin{aligned}
    p_i
    &=\mathrm{Softmax}(fc(F_{\mathrm{main}}(r_i)));\\
    \widetilde{p_i}
    &=\mathrm{Softmax}(dis(F_{\mathrm{aux}}(r_i))),
\end{aligned}
\end{equation}

where $\mathrm{Softmax}(\cdot)$ is a normalization function used to scale the probability values $p_i$ and $\widetilde{p_i}$ between 0 and 1. The optimization goal of the first stage is to minimize the loss function $\mathcal{L}_{\mathrm{mt}}$ presented in Equation (3), which comprises the cross-entropy loss for the recognition task $\mathcal{L}_{\mathrm{recog}}$ and the cross-entropy loss for the auxiliary task $\mathcal{L}_{\mathrm{aux}}$:

\begin{equation}
\begin{aligned}
    \mathcal{L}_{\mathrm{mt}}
    &=\mathcal{L}_{\mathrm{recog}} + \mathcal{L}_{\mathrm{aux}} =-\frac{1}{n} \sum_{i=1}^{n} (y_i \log(p_i))-\frac{1}{n} \sum_{i=1}^{n} (\widetilde{y_i} \log(\widetilde{p_i})).
\end{aligned}
\end{equation}

The loss function $\mathcal{L}_{\mathrm{mt}}$ gets minimized when the model can accurately predict the target category and estimate the influential factor simultaneously. The parameters of all layers are updated based on the gradient that minimizes $\mathcal{L}_{\mathrm{mt}}$. In the multi-task learning stage, there is a mutual learning process that fosters the complementary relationship between the two tasks. The auxiliary task can transfer implicit information regarding influential factors to the recognition task through the shared layer, allowing the recognition task to perceive influential factors. Meanwhile, the recognition task can also communicate information about target characteristic patterns to the auxiliary task. This may benefit the auxiliary task as it provides insights into intra-class differences and inter-class commonalities.

In the second training stage, known as the adversarial learning stage (see Fig.~\ref{fig:train} (b)), the recognition branch is intentionally deactivated and does not participate in the model training and parameter updating. The training process for the auxiliary branch remains consistent with that of the first training stage, but its training objective shifts to making the auxiliary branch unable to distinguish influential factors. To achieve this objective, we dynamically generate random auxiliary labels $\mathring{Y}=\{\mathring{y_1},\mathring{y_2},...,\mathring{y_i}\}$ for influential factors. For instance, if the actual water column depth of a sample is less than 6~meters (the true auxiliary label should be ``0''), we deliberately set a random auxiliary label from the label set as a misleading label (the misleading label can be ``0'' or ``1'' or ``2''). The training objective of the second stage is to make the model's prediction probability $\widetilde{p_i}$ of the influential factor close to misleading label $\mathring{y_i}$. It can be formalized as minimizing the cross-entropy loss $\mathcal{L}_{\mathrm{adv}}$: 

\begin{equation}
\begin{aligned}
    \mathcal{L}_{\mathrm{adv}}
    &=-\frac{1}{n} \sum_{i=1}^{n} (\mathring{y_i} \log(\widetilde{p_i})).
\end{aligned}
\end{equation}

The loss function $\mathcal{L}_{\mathrm{adv}}$ is minimized when the model is misled to the point where it cannot distinguish influential factors. The parameters of the auxiliary branch and the shared layer are updated according to the gradient that minimizes $\mathcal{L}_{\mathrm{adv}}$. In the adversarial stage, the model is constrained to extract indiscernible representations that are insensitive to various influential factors.

During the training stage, the model undergoes iterative epochs of these two stages until convergence is achieved. To preserve the predictive capacity of the model, adversarial training would not be performed during the final training epoch. The alternating execution of the two stages allows for the enhancement of both the model's recognition capabilities and its robustness against influential factors. From the perspective of the shared layer, multi-task learning helps in acquiring discriminative representations, whereas adversarial learning promotes the extraction of robust representations that are less susceptible to influential factors. As training progresses, the shared layer gradually develops higher-level representations that strike a balance between preserving discriminative power for target categories and reducing sensitivity to disturbances caused by influential factors.

After introducing the training process, we proceed to present the testing or inference process of AMTNet, as depicted in Fig.~\ref{fig:train} (c). The input to be predicted $x_i$ is first fed into the shared layer to derive the representation $r_i$. Then, $r_i$ is simultaneously fed into the recognition branch and the auxiliary branch, obtaining prediction probabilities $p_i$ and $\widetilde{p_i}$, respectively. The final prediction is determined by the label that corresponds to the highest prediction probability value. In addition, during the test stage, the model can optionally prune the auxiliary branch (remove the part in the dotted box in Fig.~\ref{fig:train} (c)) according to the lightweight requirements of model deployment or application scenarios. In this case, the function to predict influential factors can be discarded, while the model's parameter numbers, memory usage, and inference time (without considering multiprocess inference) can be approximately halved.

\subsection{Model Architecture}

\begin{table*}[ht]
\normalsize
    \centering
    \caption{The specific model architectures for AMTNet, along with the corresponding output feature dimension for each module. In this case, the CQT spectrogram serves as the sample input, which is extracted from the 30-second signal and has a dimension of (1,1200,399).}
	\scalebox{0.65}{\begin{tabular}{lll}
		\hline
		  Module&Specific network layer&Feature dimension\\
            \hline
            Input (signal waveform)&-& 1,1323000 (44100 Hz$\times$30 seconds)\\
            \hline
            Feature extraction&-&1,1200,399 (CQT spec)\\
            \hline
            Shared layer&Conv2d(1, 64, kernel size=7, stride=2, padding=3)&64,600,200 \\
            &Batch Normalization 2d(num features=64)&64,600,200\\ 
             &ReLU()&64,600,200\\ 
             &Max pooling(kernel size=3, stride=2, padding=1)&64,300,100\\
             \hline
             Main body of recognition branch&Basic block$^{\star}$(64,64),Basic block(64,64)&64,300,100\\
             \&auxiliary branch&Basic block(64,128),Basic block(128,128)&128,150,50\\
             &Basic block(128,256), Basic block(256,256)&256,75,25\\
             &Basic block(256,512), Basic block(512,512)&512,38,13\\
             &Adaptive average pooling(output size=(1, 1))&512, 1, 1\\
            \hline  % here
            %\hdashline
            $\star$Basic block(in dim, out dim)& Conv2d(in dim, out dim, kernel size=3, padding=1)\\
            & Batch Normalization 2d(num features=out dim)\\
            &ReLU()\\ 
            & Conv2d(out dim, out dim, kernel size=3, padding=1)\\
            & Batch Normalization 2d(num features=out dim)\\
		\hline
            Fully connected layer (recognition branch)&Linear(in features=512, out features=$n_{\mathrm{class}}$)&$n_{\mathrm{class}}$ ($n_{\mathrm{class
            }}$=12)\\
            \hline
            Discriminator (auxiliary branch)&Linear(in features=512, out features=$n_{\mathrm{aux}}$)&$n_{\mathrm{aux}}$ ($n_{\mathrm{aux}}$=2 or 3)\\
            \hline
        \label{tab_structure}
	\end{tabular}}
\end{table*}

The specific architecture of our proposed AMTNet is illustrated in Table~\ref{tab_structure}. At the front end, the model comprises a shared layer that incorporates a convolution layer, a batch normalization (BN) layer, a ReLU layer, and a max-pooling layer. Following the shared layer, the model diverges into two distinct branches: the recognition branch and the auxiliary branch. The main body of both branches follows the structure of four residual layers adopted in ResNet-18~\cite{he2016deep}. Each residual layer consists of two basic blocks, each containing two convolution layers, two BN layers, a ReLU layer, and a skip connection (refer to Fig.~\ref{fig:train} (d) and Table~\ref{tab_structure}$\star$). Subsequently, an average pooling layer is utilized, along with a flattening operation, to convert the feature map into a flattened vector. In the recognition branch, the flattened vector is passed through a fully connected layer, which is a simple linear layer, to transform it into the predicted probability. Regarding the auxiliary branch, its output layer, called discriminator, is also a linear layer but with different output dimensions depending on the specific task. To ensure reproducibility, the output dimensions of each layer are provided in the rightmost column of Table~\ref{tab_structure}.

\subsection{Optimization Techniques for AMT}

This subsection delves into two customized optimization techniques to address potential issues of our proposed adversarial multi-task model. First, we observed that each iteration epoch, except for the final one, ends after executing adversarial learning. At this point, not only does the auxiliary task lose its discriminative capacity, but the fluctuations in the shared layer's parameters also disrupt the intended functionality of the recognition branch. This consequently leads to poor model performance during validation at the end of each iteration epoch, thereby rendering the selection of the optimal model unattainable. To mitigate this issue, we reversed the sequence of the two stages. This adjustment guarantees that each training iteration ends with multi-task learning. This modification ensures that both tasks can achieve reliable performance during validation after each training iteration.

Moreover, we consider it crucial to slow down the parameter updates during the adversarial learning stage. If both stages have the same rate for parameter updating, the parameter updates for the shared layer and the auxiliary branch would oscillate back and forth, making it difficult to optimize towards our desired direction~\cite{liu2017adversarial,mao2020adaptive,liang2020aspect}. Such a scenario poses the risk of the model struggling to converge. To mitigate this issue, we drew inspiration from previous research on adversarial multi-task learning, wherein lower loss or learning rates were set for the adversarial stage~\cite{liu2017adversarial}. In this study, we assign a lower learning rate for the adversarial learning stage, which is one-fifth of the learning rate used in the multi-task learning stage. This adjustment encourages the shared layer to prioritize learning high-level representations that are less susceptible to influential factors, rather than relying on drastic parameter changes to achieve adversarial regularization. In Section V.C. and Table~\ref{tab:ablation}, we conducted ablation experiments to verify the necessity of these two optimization techniques.

% the shared layers failing to learn high-level representations. This occurs because the shared layers may prioritize disrupting the auxiliary branches by making significant parameter changes, \textcolor{red}{rather than focusing on learning meaningful representations.}

% \footnote{The dataset is available at http://atlanttic.uvigo.es/underwaternoise.}

\section{Experimental Setup}
\subsection{Dataset}
In this study, we conducted all experiments on an underwater ship-radiated noise dataset - ShipsEar~\cite{santos2016shipsear}. ShipsEar is an open-source database of underwater recordings of ship-radiated sounds. The recordings were collected from various locations along the Atlantic coast of Spain between 2012 and 2014, using single or multiple hydrophones. The dataset consists of 90 recordings, sampled at a rate of 52,734 Hz, with durations ranging from 15 seconds to 10 minutes. The cumulative duration of all recordings is approximately three hours. It includes 11 distinct types of ship sounds and 1 type of natural noise, including dredgers, fish boats, motorboats, mussel boats, ocean liners, passenger ships, ro-ro ships, sailboats, pilot ships, trawlers, tug boats, and natural noise (the recognition task on ShipsEar can be regarded as a 12-class classification task). In addition to the signal recordings and corresponding category labels, ShipsEar also includes annotations for available metadata, which were recorded during data acquisition. These annotations include the approximate source range, water column depth, wind speed measured in situ, the GPS position of the recording equipment, time and date of the recording, gains and depths of the hydrophones, and atmospheric and oceanographic data. In this study, we carefully selected three factors for investigation, taking into account the completeness of the annotations and the extent to which they influence underwater signals.

\subsection{Data division}
In this study, we divided each signal recording into consecutive segments of 30 seconds, with a 15-second overlap. To mitigate the risk of information leakage, we ensured that the segments in the training and test sets originated from distinct signal recordings. By imposing such a rigorous restriction, we aim to ensure that the obtained test results can reasonably reflect the model's generalization performance to unseen targets. Moreover, the absence of a standardized train-test split in ShipsEar introduces challenges when comparing results across different studies. To address this issue and promote fair comparisons, we have provided our own train-test split, as detailed in Table~\ref{tab:split}. Our purpose is to establish a reliable benchmark that fosters fairness in comparative evaluations. The specific train-test split, accompanied by the corresponding code, is publicly available at: \url{https://github.com/xy980523/ShipsEar-An-Unofficial-Train-Test-Split}.

% 15\% of the data in the training set is randomly chosen as the validation set.
\begin{table*}[ht]
\normalsize
    \caption{\label{tab:split} The train-test split for ShipsEar. The ``ID'' column corresponds to the identification records of the ``.wav'' files in the dataset. The ``Train/test records num'' column indicates the number of complete recordings included in the training and test sets, while the ``Train/test segments num'' column indicates the number of 30-second segments in each set. Notably, for ``Trawler'', which consists of a single recording, this study divides the recording into two non-overlapping segments, which are individually used as training and test samples.}
    \centering
	\scalebox{0.7}{
	\begin{tabular}{lllcc}
        \hline
	Category&Training data ID &Test data ID & Train/test records num&  Train/test segments num\\
	\hline
	Dredger&80,94,96& 93,95&3/2 &8/4\\
    Fish boat&73,74,76&75&3/1 &26/2\\
    Motorboat&27,33,39,45,50,52,70,72,77,79&21,26,51&10/3 &43/6\\
    Mussel boat&47,48,49&46,66 &3/2&33/9\\
    Natural noise&82,83,84,86,87,89,90,91,92&81,85,88&9/3&45/16\\
    Ocean liner&16,23,25,69,71&22,24&5/2&50/3\\
    Passenger ship&06,08,09,10,11,12,13,17,32,34,36,38,&07,14,35,40,43,54,61&23/7&217/25\\
    &41,42,53,55,59,60,62,63,64,65,67&&& \\
    Pilotship&29 &30&1/1 &5/1\\
    RO-RO ship&18,19,58&20,78&3/2&81/12\\
    Sailboat&37,57,68&56&3/1&19/2\\
    Trawler&28 (15-125 second) & 28 (125-163 second)&1/1&6/1\\
    Tugboat&15 & 31&1/1&8/3\\
    \hline
    Total Number& - &-&65/26 &541/84 \\
    Train-test percent&-&-&71.43\%/28.57\%&-\\
    \hline
	\end{tabular}}
\end{table*}

\subsection{Parameters Setup}
The parameter setup for preprocessing and feature extraction can be found in Section III.A. For training, we utilize the AdamW optimizer~\cite{loshchilov2017decoupled}. The maximum learning rate is set to 5$\times 10^{-4}$ for the multi-task learning stage, and 1$\times 10^{-4}$ for the adversarial learning stage. We employed the cosine annealing algorithm~\cite{loshchilov2016sgdr} to schedule the learning rate decay. All models are trained for 200 epochs with a warm-up epoch of 5 and a weight decay of $10^{-5}$. The experiments were conducted on Nvidia A10 GPUs, with CUDA version 11.4. The Python version used is 3.8.8 and the Pytorch version is 1.9.0.

\section{Results and Analyses}

This section initiates with preliminary experiments to determine the optimal filter cutoff frequency, backbone model architecture, and acoustic feature in subsection A. Then, we introduce a series of comparative and ablation experiments to demonstrate the superiority of our proposed AMTNet and optimization techniques in subsections B and C. Additionally, we provide in-depth analyses to elucidate the effectiveness of AMTNet in obtaining robust representations in subsection D. Finally, we evaluate the performance of auxiliary tasks and assess their potential impact on recognition in subsection E.

In this study, we employed accuracy as the uniform evaluation metric for all experiments, calculated by dividing the number of correctly predicted samples by the total number of samples. Notably, we reported accuracy at the 30-second segment level instead of the record-file level to avoid the occurrence of identical file-level accuracy among multiple groups of experiments due to the limited number of record files in the test set. To mitigate the impact of randomness, each experiment was performed twice using different random seeds (123 and 3407), and we reported the mean and standard deviation (stdev) of the segment-level accuracy produced by the two runs. For instance, we would report 75.60(mean)$\pm$0.42(stdev)\% when the accuracy of two runs with distinct random seeds was 75.00\% and 76.19\%. Furthermore, during the training process, we uniformly employed a data augmentation technique known as local masking and replicating (LMR)~\cite{xu2023underwater}. This approach generates mixed spectrograms by randomly masking local patches in passbands and replicating patches from the same region in other spectrograms, which has proven beneficial in recognizing underwater acoustic signals based on time-frequency spectrograms.

\subsection{Preliminary Experiments}

\begin{figure*}
    \centering
    \includegraphics[width=0.9\linewidth]{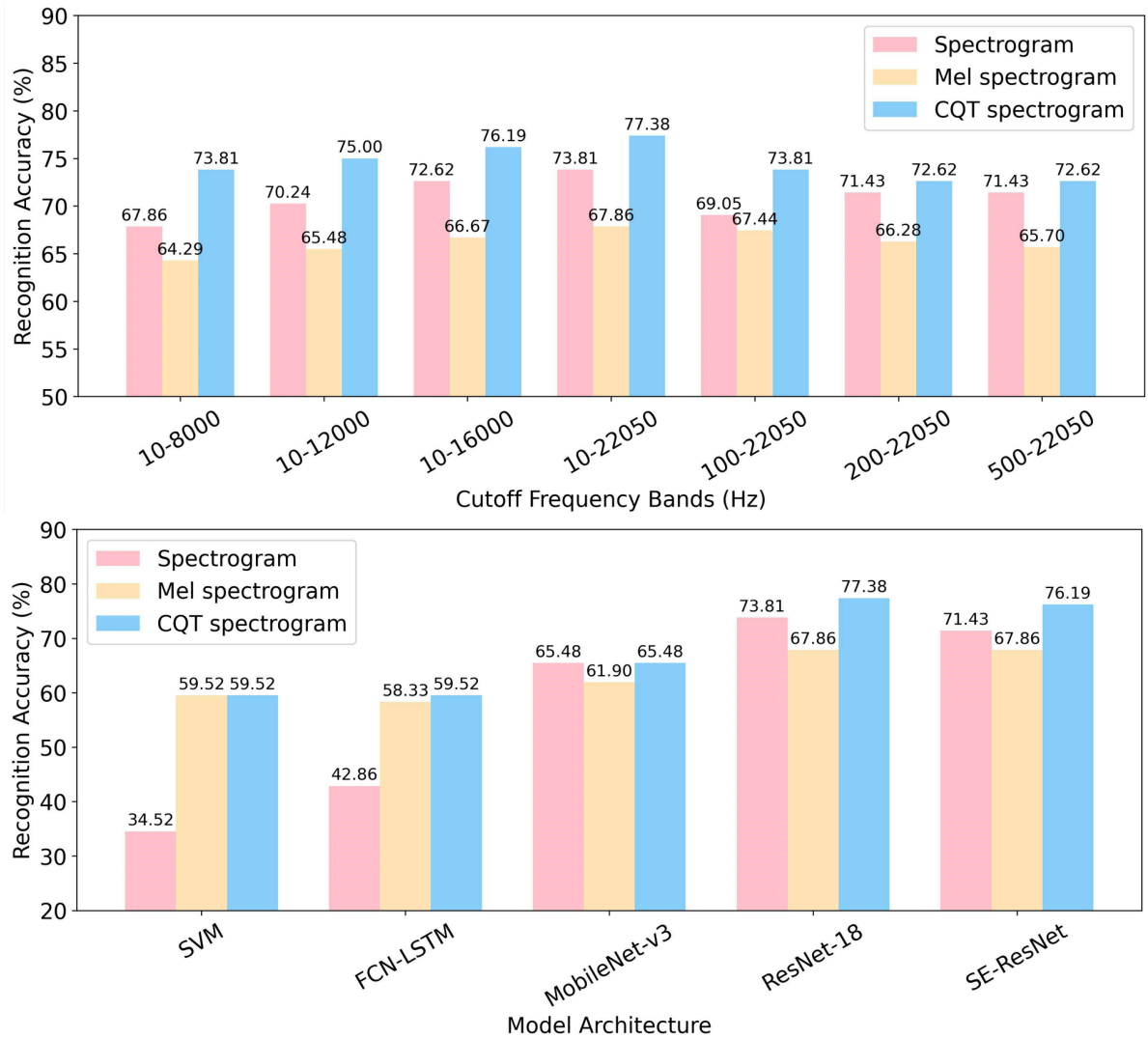}
    \caption{Preliminary experiments on selecting the optimal filter cutoff frequency, backbone model architecture, and acoustic feature.}
    \label{fig:pre}
    \vspace{-2px}
\end{figure*}

Before validating our proposed approaches, we performed preliminary experiments to determine the optimal filter cutoff frequency, backbone model architecture, and acoustic feature. The detailed results are presented in Fig.~\ref{fig:pre}. Firstly, we conducted comparative experiments on various cutoff frequency bands using three input features, with ResNet-18 as the default backbone model. We considered several candidate frequency values for testing, including lower frequencies of 10~Hz, 100~Hz, 200~Hz, and 500~Hz, as well as upper frequencies of 8000~Hz, 12000~Hz, 16000~Hz, and 22050~Hz. As shown in the upper part of Fig.~\ref{fig:pre}, the frequency band of 10-22050Hz consistently achieves the best recognition performance across the three input features. Therefore, we set the default filter cutoff frequency to 10-22050~Hz.

Moreover, we evaluated the performance of different backbone models using various input features. The candidate models included support vector machine (SVM)~\cite{hearst1998support}, fully convolutional network-long short-term memory (FCN-LSTM)~\cite{shi2015convolutional}, MobileNet-v3 small~\cite{howard2019searching}, Residual Network-18 (ResNet-18)~\cite{he2016deep}, Squeeze-and-Excitation Residual Network (SE-ResNet)~\cite{hu2018squeeze}. As shown in the lower portion of Fig.~\ref{fig:pre}, the results indicate that the vanilla ResNet-18 outperformed other backbone models, even surpassing its optimized variants, such as SE-ResNet. This observation suggests that the structure and complexity of ResNet-18 are well-suited for the task and can serve as a reliable backbone model. In this study, AMTNet adopts the structure of ResNet-18 for its various modules, where the shared layer conforms to the front-end layers of ResNet-18, and the main body of each branch exactly follows the four residual layers of ResNet-18.

Furthermore, the preliminary experiments also validate the performance of three candidate features. As shown in Fig.~\ref{fig:pre}, except when utilizing SVM or FCN-LSTM as the backbone model, the models based on Mel spectrograms exhibit inferior performance compared to those based on other features. This disparity can be attributed to the susceptibility of low-frequency signal components to interference from distant sources, while the higher low-frequency frequency resolution of the Mel spectrogram may introduce more target-irrelevant noise components, thereby compromising performance. In contrast, CQT spectrograms offer high temporal resolution in the high-frequency range, which facilitates the reflection of periodic rhythm information in signals, such as the propeller beat, thus resulting in superior performance compared to other features. Based on these findings, we designated the CQT spectrogram as the default input feature for this study.

\subsection{Comparative Experiments}

\begin{table*}[ht]
\normalsize
    \caption{\label{tab:main} The comparative experiments on ShipsEar, which include our reproduced advanced methods as benchmarks. The ``Influential factor'' notation indicates whether and which influential factors are utilized.}
    \centering
	\scalebox{0.75}{
	\begin{tabular}{llcc}
        \hline
	Feature&Model&Influential factor&12-class accuracy(\%)\\
	\hline
	\textbf{Advanced methodology
} & \textbf{as benchmarks}&&\\
        Wavelet spec & AGNet~\cite{xie2022adaptive} & $\times$& 72.62$\pm$0.84 \\
        Gabor spec & UALF~\cite{ren2022ualf} & $\times$& 71.43$\pm$1.27\\
        CQT spec & SIR\&LMR~\cite{xu2023underwater} & $\times$& 77.38$\pm$0.42\\
        CQT spec\&Mel spec& ICL~\cite{xie2023guiding} & $\times$& 77.98$\pm$1.27\\
        CQT spec\&STFT spec & ICL & $\times$& 79.76$\pm$0.00\\
        Wavelet spec & UART~\cite{xie2022underwater} & $\checkmark$ (source range)& 75.00$\pm$0.00 \\
          &   & $\checkmark$ (water column depth)& 72.62$\pm$0.84 \\
          &   & $\checkmark$ (wind speed)& 72.62$\pm$0.00 \\
        \hline
        \textbf{Our approaches} &&&\\
        CQT spec & ResNet-18 (baseline)& $\times$&77.38$\pm$0.42\\
        CQT spec& MTNet (AMTNet w/o adversarial learning) &$\checkmark$ (source range)&77.98$\pm$0.42 \\
        &&$\checkmark$ (water column depth)&74.40$\pm$0.42 \\
        &&$\checkmark$ (wind speed)&75.60$\pm$0.42 \\
        
        CQT spec& AMTNet &$\checkmark$ (source range)&80.95$\pm$0.84 \\
        &&$\checkmark$ (water column depth)&79.76$\pm$0.84 \\
        &&$\checkmark$ (wind speed)&80.36$\pm$0.42 \\
        \hline
	\end{tabular}}
\end{table*}

After conducting preliminary experiments to determine the default configurations, we proceeded with comparative experiments to validate the superiority of AMTNet over current advanced methods. The detailed results of the comparative experiments are illustrated in Table~\ref{tab:main}. To ensure a fair comparison, we unified the train-test-split and replicated several advanced works, including AGNet~\cite{xie2022adaptive}, UALF~\cite{ren2022ualf}, SIR+LMR~\cite{xu2023underwater}, ICL~\cite{xie2023guiding}, and UART~\cite{xie2022underwater}, on ShipsEar as benchmarks. Among these, ICL achieved the best performance with an accuracy of 79.76\%. Besides, we observed that UART, which also models influential factors in ShipsEar, failed to deliver satisfactory results. Although this could be attributed to the incompatibility between the UART's feature and model with the 12-class recognition task, it implies that existing methods have significant room for improvement in modeling influential factors. In comparison to existing methods, our approaches can achieve satisfactory results. The optimal baseline model, which was selected through preliminary experiments, attains a recognition accuracy of 77.38$\pm$0.42\%, surpassing several previous advanced methods. Notably, while our baseline method demonstrates competitiveness, it does not necessarily imply that our baseline method is better than previous methods with inferior results, as the experimental setups (feature, model architecture, parameter configuration) of those previous methods may not be directly applicable to the 12-class recognition task, resulting in a significant performance discount due to the differences in task requirements and experimental setups.

In addition, we presented the results of AMTNet without adversarial learning, which can be regarded as a vanilla multi-task model (referred to as MTNet). Our experimental findings demonstrate that performing multi-task learning without adversarial training may lead to unsatisfactory recognition performance. Specifically, when estimating water column depth or wind speed was employed as the auxiliary task, the recognition accuracy of the multi-task model significantly decreased compared to the baseline model. This discrepancy arises due to the mismatch between the objectives of the auxiliary task and the recognition task, causing mutual interference and negative transfer phenomenon~\cite{tang2020progressive}. By incorporating adversarial learning, AMTNet's recognition performance can be substantially enhanced across the three influential factors. It can bring about an evident recognition accuracy improvement of around 2.38\% $\sim$ 3.57\% compared to the baseline and achieve state-of-the-art performance (80.95$\pm$0.84\%) on the 12-class recognition task in ShipsEar. Notably, the performance gain achieved through adversarial training is particularly pronounced. Especially for models based on water column depth or wind speed, the incorporation of adversarial learning can bring approximately 5\% performance improvements. This notable improvement can be attributed to the fact that adversarial learning not only enables the shared layer to acquire robust representations that are insensitive to influential factors but also prevents the shared layer from capturing discriminative patterns related to influential factors. When confronted with the mutual interference between tasks, the model tends to prioritize patterns relevant to the target category, thereby exhibiting a stronger inclination towards the recognition task. In other words, adversarial learning does not resolve the issue of inter-task interference, but instead prioritizes the recognition task at the expense of slightly compromising the performance of the auxiliary task.

% Additionally, AMTNet based on source range and water column depth can also lead to accuracy improvements of 1.79\% and 2.38\%, respectively, when compared to the baseline. The superiority of the AMT-based models over the inferior models based solely on multi-task learning confirms the necessity of the adversarial learning stage. The adversarial discriminative paradigm compels the shared layer to learn high-level representations that are insensitive to condition variations. This insensitivity not only aids in reducing the mutual interference between the recognition task and the auxiliary task, but also enhances the model's robustness and generalization capabilities. Further statistical analyses of the enhanced model's robustness will be provided in subsection D.

\subsection{Ablation Experiments on Optimization Techniques}
\begin{table*}[ht]
\normalsize
    \caption{\label{tab:ablation} The ablation experiments on optimization techniques of AMTNet. ``Iteration reversal'' represents reversing the iteration sequence of the two training stages, and ``Less learning rate (1/n)'' represents setting a lower learning rate for the adversarial learning stage, which is 1/n of the learning rate for the multi-task learning stage.}
    \centering
	\scalebox{0.9}{
	\begin{tabular}{ccccc}
        \hline
	Iteration reversal&\quad Less learning rate (1/10)\quad&\quad Less learning rate (1/5) \quad&12-class accuracy(\%)&Benefits(\%)\\
        \hline
        \textbf{source range} &&&&\\
	$\times$ & $\times$ &$\times$ &               75.60$\pm$0.42 & -\\
        $\checkmark$ & $\times$ &$\times$ &
        76.79$\pm$0.42 &+1.19\\
        $\times$ & $\checkmark$ &$\times$ & 77.98$\pm$0.42&+2.38\\
        $\times$ & $\times$ &$\checkmark$ &
        78.57$\pm$0.00 &+2.97\\
        $\checkmark$ & $\checkmark$ &$\checkmark$ &80.95$\pm$0.84 &+5.35\\
        \hline
        \textbf{water column depth} &&&&\\
	$\times$ & $\times$ &$\times$ &               76.79$\pm$0.42 & -\\
        $\checkmark$ & $\times$ &$\times$ &
        77.38$\pm$0.00 &+0.59\\
        $\times$ & $\checkmark$ &$\times$ &
        77.98$\pm$0.00 &+1.19\\
        $\times$ & $\times$ &$\checkmark$ &
        77.98$\pm$0.42 &+1.19\\
        $\checkmark$ & $\checkmark$ &$\checkmark$ &79.76$\pm$0.84 &+2.97\\
        \hline
        \textbf{Wind speed} &&\\
	$\times$ & $\times$ &$\times$ &
        76.19$\pm$0.00 & -\\
        $\checkmark$ & $\times$ &$\times$ &
        76.79$\pm$0.42 &+0.60\\
        $\times$ & $\checkmark$ &$\times$ & 77.38$\pm$1.68&+1.19\\
        $\times$ & $\times$ &$\checkmark$ & 77.98$\pm$2.11&+1.79\\
        $\checkmark$ & $\checkmark$ &$\checkmark$ &80.36$\pm$0.42 &+4.17\\
        \hline
	\end{tabular}}
\end{table*}

In addition, we conducted ablation experiments to verify the efficacy of our proposed two optimization techniques and presented the results in Table~\ref{tab:ablation}. Firstly, the iteration reversal, which represents reversing the iteration sequence of the two training stages, leads to a performance improvement of approximately 0.59\% $\sim$ 1.19\%. This observation demonstrates that a well-organized arrangement of training stages consistently yields benefits. Moreover, the results indicate that setting a lower learning rate for adversarial learning can also contribute to performance gains. By setting a 1/5 learning rate ($1\times10^{-4}$) for adversarial learning, the recognition accuracy can improve about 1.19\% $\sim$ 2.97\% compared to the vanilla AMTNet. This implies that reducing the rate of parameter updates during adversarial learning helps prevent oscillation in the optimization process of shared layer parameters. By implementing both optimization techniques simultaneously, the performance of AMTNet experienced an additional enhancement. Particularly for the AMTNet utilizing ``source range'' as the influential factor, the two optimization techniques can result in a remarkable improvement of 5.35\%, bringing the recognition accuracy to the state-of-the-art level.

%Regardless of the environmental conditions used in the AMT-based model, the improvement brought by applying two optimization techniques at the same time is greater than the sum of the improvements brought by applying the optimization strategy separately. This shows that the two optimization techniques do not affect each other and are complementary to each other.

Besides, we have observed a noticeable decrease in the recognition performance of the vanilla AMTNet (75.60\% $\sim$ 76.79\%) compared to the baseline (77.38\%). It indicates that the vanilla AMTNet is affected by underlying issues that impact its recognition performance. These issues may include the mutual influence between tasks or the parametric oscillation and performance drop caused by excessive adversarial learning. Whereas, by incorporating our proposed optimization techniques, some of these underlying issues can be effectively alleviated, allowing the potential of AMTNet to be realized and leading to a remarkable performance breakthrough. Furthermore, we discovered that the simultaneous application of both optimization techniques results in greater gains compared to applying each technique individually. This finding suggests that our proposed optimization strategies exhibit complementarity effects. For instance, reversing the iteration sequence enables the model to achieve accurate recognition results by the end of iterations. However, placing adversarial learning at the forefront can lead to unstable parameter optimization directions during the initial training stages. In such situations, setting a lower learning rate for adversarial learning can regulate the rate of parameter updates. This adjustment ensures that the optimization direction remains favorable for recognition and helps compensate for potential drawbacks caused by reversing the iteration sequence.

\subsection{Additional Analyses on Robustness}

\begin{figure*}
    \centering
    \includegraphics[width=1.0\linewidth]{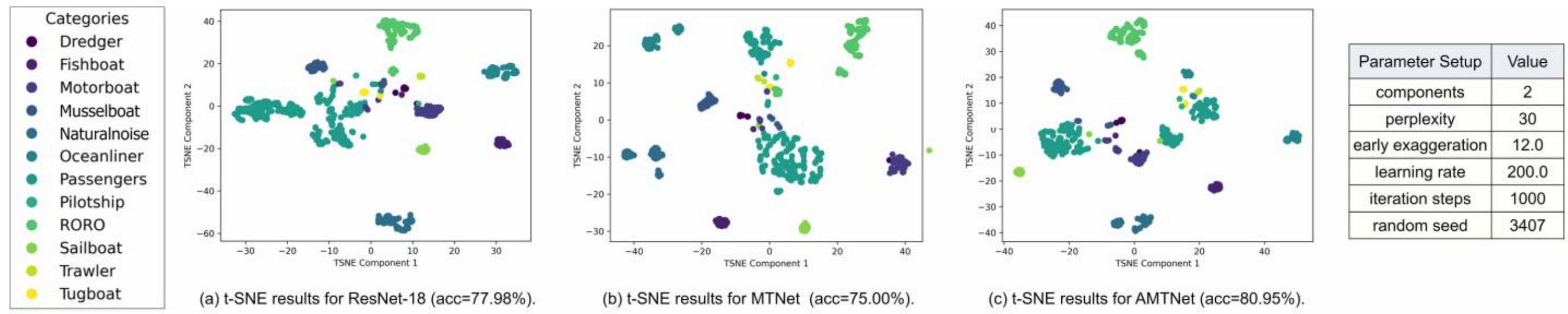}
    \caption{The t-SNE results and corresponding details. The MTNet and AMTNet used here are unified with ``wind speed estimation'' as the auxiliary task.}
    \label{fig:cluster}
    \vspace{-2px}
\end{figure*}

After validating the recognition performance, we conducted additional analyses to evaluate the capability of AMTNet to acquire robust representations. First, we employed t-Distributed Stochastic Neighbor Embedding (t-SNE)~\cite{van2008visualizing} to visualize the distribution of intermediate layer representations from three candidate models (ResNet-18, MTNet, and AMTNet) in a low-dimensional space. This technique operates by computing pairwise probability distributions that represent similarities between data points in high-dimensional space and then uses a Student-t distribution to calculate this distribution in low-dimensional space. By minimizing the Kullback-Leibler divergence between these two distributions using a gradient descent optimization, t-SNE guarantees that similar data points maintain proximity to each other in the low-dimensional space, enabling the visualization of the underlying data structure in low-dimensional space. In this study, the output of the shared layer served as the intermediate layer representation (for ResNet-18 without the shared layer, the output of the max-pooling layer served as the intermediate layer representation). Fig.~\ref{fig:cluster} illustrates the t-SNE results of the intermediate layer representations of the three models on the entire ShipsEar dataset, with the corresponding parameter setups listed on the right side of the figure. It is evident from Fig.~\ref{fig:cluster} (c) that the data points based on AMTNet exhibit smaller intra-class distances, indicating that the intermediate layer representations of the same category can maintain similar distributions under different influential factors. This demonstrates that AMTNet effectively extracts robust representations that are insensitive to influential factors. Moreover, smaller intra-class distances can also contribute to improved recognition performance~\cite{schutz2013k}, as they facilitate better decision boundary delineation.

% To further strengthen the argument, we also report the cluster inertia in Fig.~\ref{fig:cluster}, which represents the sum of distances between all samples and their respective cluster centroids. A smaller cluster inertia value indicates less intra-class diversity among representations. Among the three models, the AMT-based model has a cluster inertia of 53208.78, significantly smaller than the 62034.46 and 81077.67 of the other two models. This quantitative analysis confirms that the AMT-based model is less susceptible to variations and exhibits enhanced robustness.

\begin{table*}[ht]
\normalsize
    \caption{\label{tab:cos_sim} The cosine similarity of the predictions for three different recordings of the Passenger ship ``Minho Uno'' based on different models. The MTNet and AMTNet used here are unified with ``wind speed estimation'' as the auxiliary task.}
    \centering
	\scalebox{0.85}{
	\begin{tabular}{lccc}
        \hline
	Recordings& \quad Similarity on ResNet-18 \quad & \quad Similarity on MTNet \quad &  \quad Similarity on AMTNet \quad \\
        \hline
        Passenger 11 \& 36 \quad &0.8324&0.9310&0.9410\\
        Passenger 11 \& 65  \quad &0.8039&0.7158&0.8675\\
        Passenger 36 \& 65 \quad &0.8329&0.8296&0.8514\\
        \hline
        Average &0.8231&0.8255&0.8866\\
        \hline
	\end{tabular}}
\end{table*}

We then proceeded to investigate whether the model can maintain this robustness during prediction. The analyses primarily revolve around three specific recordings in ShipsEar: Passenger 11, Passenger 36, and Passenger 65, which have been introduced in Fig.~\ref{fig:background}. These recordings belong to the same target (passenger ship ``Minho Uno''), with the only distinction being their influential factors. We uniformly extracted the initial 30 seconds from these recordings and fed them into the model, resulting in the retrieval of embedding vectors $E_{11},E_{36},E_{65}\sim \mathcal{R}^{1\times512}$ (the subscript ``11,36,65'' denotes the recording ID) before the last linear layer. Then, these embedding vectors were pairwise crossed to calculate the cosine similarity, which signifies the similarity of the model's predictions for the corresponding recordings. The cosine similarity can be calculated using the following formula:

\begin{equation}
   \text{Similarity}(E_A, E_B) = \frac{E_A \cdot E_B^{T}}{{\|E_A\| \cdot \|E_B\|}}.
\end{equation}

We reported the cosine similarity of the embedding vectors in Table~\ref{tab:cos_sim}. AMTNet can exhibit an average cosine similarity of 0.8866 across the three recordings, surpassing the other models (0.8231 and 0.8255). This suggests that AMTNet can make more similar predictions for signals belonging to the same target but under different influential factors. This further demonstrates that AMTNet prioritizes the intrinsic characteristics of targets while minimizing the undesired association of target-specific patterns with influential factors.

\subsection{Results of Auxiliary Tasks}

% \begin{figure}
%     \centering
%     \includegraphics[width=1.0\linewidth]{Figure5.eps}
%     \caption{The recognition accuracy of the multi-task model and AMT-based model on three auxiliary tasks: source range estimation, water column depth estimation, and wind speed estimation.}
%     \label{fig:aux_results}
%     \vspace{-2px}
% \end{figure}

\begin{table*}[htbp]
\normalsize
    \caption{\label{tab:auxresuly} The recognition accuracy on three auxiliary tasks: source range estimation, water column depth estimation, and wind speed estimation.}
    \centering
	\scalebox{0.9}{
	\begin{tabular}{lcc}
        \hline
	Task&Acc on MTNet (\%)& Acc on AMTNet(\%)\\
        \hline
        Source range estimation (2-class classification)&83.93$\pm$1.27 &82.74$\pm$1.27\\
        Water column depth estimation (3-class classification)&90.48$\pm$4.21&91.08$\pm$2.11 \\
        Wind speed estimation (3-class classification)&88.70$\pm$0.42 &87.21$\pm$1.47\\
	\hline

        \hline
	\end{tabular}}
\end{table*}

In this subsection, we presented the results of auxiliary tasks concerning influential factors, including source range estimation, water column depth estimation, and wind speed estimation. Table~\ref{tab:auxresuly} illustrates the recognition accuracy of MTNet and AMTNet on the three auxiliary tasks. The results demonstrate that AMTNet and MTNet can exhibit comparable performance. For water column depth estimation, AMTNet can even outperform MTNet. This is a somewhat unexpected result, as the objective of adversarial learning ostensibly compromises the performance of auxiliary tasks. We can only attribute this to the robustness and resilience afforded by adversarial learning can sometimes outweigh its potential negative effects. In general, AMTNet has demonstrated its efficacy in auxiliary tasks. In certain application scenarios, it is capable of providing a rough estimation of environmental conditions or data acquisition configurations.

% The results demonstrate that both models perform poorly in source range estimation, achieving a maximum accuracy of only 76.19\% in the 3-class auxiliary task. However, in the tasks of estimating water column depth and wind speed, both models exhibit satisfactory performance. Particularly, the multi-task model can achieve a maximum accuracy of 96.44\% in the water column depth estimation. Furthermore, we observe that although the AMT-based model far surpasses the multi-task model in the recognition task, their performance in auxiliary tasks is similar, with the multi-task model often proving to be better. This phenomenon can be attributed to the inherent nature of adversarial learning, which, on the one hand, bolsters the robustness of the shared layer, but on the other hand, entails the potential risk of compromising the performance of auxiliary tasks. Fortunately, our experimental findings suggest that the incorporation of adversarial learning does not substantially impair the performance of the auxiliary tasks.

In addition, we discovered that AMTNet exhibits mediocre performance in source range estimation, achieving only approximately 82.74 $\sim$ 83.93\% accuracy in the 2-class classification task. Regarding the other two tasks, AMTNet demonstrates relatively superior performance, particularly in water column depth estimation, where it can achieve a 3-class recognition accuracy of 91.08$\pm$2.11\%. This discrepancy suggests that different influential factors possess distinct patterns and discriminability. The discriminative patterns related to source range are relatively fewer, partly due to the normalization operation during preprocessing, which eliminated the differences in sound intensity, and partly because the source range used in this study is small (0$\sim$350~meters), resulting in minimal signal distortion caused by underwater propagation. On the other hand, targets with different water column depths are more easily distinguishable, possibly due to the relatively shallow water column depths (0$\sim$20~meters) employed in this study, which amplifies the influence of seabed scattering on the signal components. Similarly, targets with different wind speeds are also more easily distinguishable, as the wind-induced surface waves (as evident in the signal waveforms in Fig.~\ref{fig:background} (b), (c)) are indeed prominent and readily distinguishable.

Furthermore, we explored the relationship between auxiliary tasks and the recognition performance of AMTNet. By comparing the results in Table~\ref{tab:main} and Table~\ref{tab:auxresuly}, we discerned a phenomenon: models that exhibit poorer performance in auxiliary tasks may achieve better performance in the recognition task. Based on our assumption, the model's poor performance in the auxiliary task indicates a lack of relevant discriminative patterns, which prompts the model to excavate more high-level patterns, thereby benefiting the recognition task. This insight provides guiding directions for future research, indicating that the design of more intricate or ingenious auxiliary tasks may further enhance the recognition performance of models.

% Furthermore, we have discovered that the performance of recognition tasks is indeed influenced to some extent by the performance of auxiliary tasks. For instance, when the source distance estimation is considered, the AMTNet model exhibits the poorest performance in both the auxiliary and recognition tasks. This can be attributed to the fact that when the auxiliary task is comparatively challenging, it leads to a higher auxiliary loss, thereby assigning greater emphasis on optimizing the auxiliary task during model training (in other words, $\mathcal{L}_{aux}$ is dominant in $\mathcal{L}_{1}$), consequently compromising the recognition task. Conversely, for tasks such as water column depth estimation and wind speed estimation, which exhibit better performance, their auxiliary loss is lower, thus exerting a reduced influence on recognition. Additionally, we hypothesize that the greater the influence of environmental conditions on underwater signals, the more discernible environment-related patterns the signals contain, resulting in relatively ``simpler'' corresponding auxiliary tasks. In this scenario, modeling this crucial environmental condition would yield greater information gain, reducing information entropy and culminating in a more comprehensive and superior model.

\section{Conclusions}

This study exposes the limitations of existing underwater acoustic target recognition methods, which are susceptible to environmental conditions or data acquisition configurations. To address this issue, we design auxiliary tasks that model influential factors based on available annotations and adopt a multi-task framework to connect these factors and the recognition task. Furthermore, we incorporate an adversarial learning mechanism into the multi-task framework to enhance the model's robustness against influential factors. With the adoption of two customized optimization techniques, our proposed AMTNet has demonstrated substantial performance improvements compared to baselines and current advanced methods, achieving state-of-the-art results on the ShipsEar dataset.

Despite AMTNet's effectiveness on ShipsEar, there are still limitations that can be addressed for further improvement. First, this study exclusively utilized the ShipsEar dataset, focusing on data collected in coastal areas with small source-receiver distances and shallow water column depths. Models trained on this dataset may generalize well to nearshore, close-range scenarios. However, their capability to generalize to offshore, long-range scenarios remains uncertain. Another challenge is that the training of AMTNet necessitates annotations for influential factors, thereby constraining our capacity to train it on other large-scale datasets lacking relevant annotations. In future work, we plan to explore the possibility of extending AMTNet into a semi-supervised learning framework. The framework would allow training with a limited amount of annotated data and large-scale unlabeled data, ultimately enhancing its generalization capability. In addition, there is room for optimization in the design of auxiliary tasks. Currently, these tasks rely on a manual label mapping standard to categorize diverse influential factors into several classes for model training. However, this design may not be optimal. In the future, with the availability of sufficient annotated data or advancements in semi-supervised learning models, we anticipate that the auxiliary task can be optimized as the regression task or other ingenious task, thus enabling the models to gain a deeper understanding of the influential factors.

% In future work, we intend to explore more sophisticated designs to minimize the risk of performance degradation when incorporating environmental conditions. Additionally, while our AMTNet currently models only one environmental condition at a time to avoid increasing model complexity, we plan to investigate techniques that can simultaneously incorporate all three environmental conditions without increasing the model complexity. By doing so, we aim to push the performance of the underwater acoustic recognition model to its upper limits

%% before appendix (optional) and bibliography:
% \begin{acknowledgments}
% This research was partially supported by the Chinese Academy of Sciences Strategic Leading Science and Technology Project (No. XDA0310103); the IOA Frontier Exploration Project (No. ZYTS202001); and the Youth Innovation Promotion Association CAS.
% \end{acknowledgments}

\section*{Author Declarations}
\textbf{Conflict of Interest} The authors declare that they have no known competing financial interests or personal relationships that could have appeared to influence the work reported in this paper.

\section*{Data Availability}
The ShipsEar dataset that supports the findings of this study is openly available at: \url{https://underwaternoise.atlanttic.uvigo.es/}; Our train-test-split is available at: \url{https://github.com/xy980523/ShipsEar-An-Unofficial-Train-Test-Split}; The partial metadata of ShipsEar, including location, date, channel depth (water column depth), wind speed, and distance (source range), is summarized at: \url{https://github.com/xy980523/ShipsEar-An-Unofficial-Train-Test-Split/blob/main/meta_info.txt}.

\newpage
% \section*{References}
% \bibliography{cas-refs}

\printcredits

\bibliographystyle{cas-model2-names}
\bibliography{cas-refs-amt}





% \end{thebibliography}
\end{document}